\documentclass{article}

\usepackage{PRIMEarxiv}

\usepackage[utf8]{inputenc} 
\usepackage[T1]{fontenc}    
\usepackage{hyperref}       
\usepackage{url}            
\usepackage{booktabs}       
\usepackage{amsfonts}       
\usepackage{nicefrac}       
\usepackage{microtype}      
\usepackage{lipsum}
\usepackage{fancyhdr}       
\usepackage{graphicx}       
\usepackage{colortbl}
\usepackage{longtable}
\usepackage{hhline}
\graphicspath{{media/}}     
\usepackage[acronym]{glossaries}
\usepackage{caption} 
\newacronym{ml}{ML}{Machine Learning}
\newacronym{ai}{AI}{Artificial Intelligence}
\newacronym{lod}{LOD}{Linked Open Data}
\newacronym{uri}{URI}{Uniform Resource Identifier}
\newacronym{rdf}{RDF}{Resource Description Framework}
\newacronym{www}{WWW}{World Wide Web}
\newacronym{kg}{KG}{Knowledge Graph}
\newacronym{sti}{STI}{Semantic Table Interpretation}
\newacronym{cea}{CEA}{Cell Entity Annotation}
\newacronym{cta}{CTA}{Column Type Annotation}
\newacronym{cpa}{CPA}{Column Predicate Annotation}

\title{SemTUI: a Framework for the Interactive Semantic Enrichment of Tabular Data}

\author{
  Marco Ripamonti \\
  University of Milan-Bicocca \\
  \texttt{m.ripamonti@campus.unimib.it} \\
  \And
  Flavio De Paoli, Matteo Palmonari \\
  University of Milan-Bicocca \\
  \texttt{\{flavio.depaoli, matteo.palmonari\}@unimib.it} \\
}

\begin{document}
\maketitle

\begin{abstract}
The large availability of datasets fosters the use of \acrshort{ml} and \acrshort{ai} technologies to gather insights, study trends, and predict unseen behaviours out of the world of data. Today, gathering and integrating data from different sources is mainly a manual activity that requires the knowledge of expert users at an high cost in terms of both time and money. It is, therefore, necessary to make the process of gathering and linking data from many different sources affordable to make datasets ready to perform the desired analysis. In this work, we propose the development of a comprehensive framework, named SemTUI, to make the enrichment process flexible, complete, and effective through the use of semantics. The approach is to promote fast integration of external services to perform enrichment tasks such as reconciliation and extension; and to provide users with a graphical interface to support additional tasks, such as refinement to correct ambiguous results provided by automatic enrichment algorithms. A task-driven user evaluation proved SemTUI to be understandable, usable, and capable of achieving table enrichment with little effort and time with user tests that involved people with different skills and experiences.
\end{abstract}


\section{Introduction}
In the last decade, the use of Data increased exponentially in all fields, from scientific research to business organizations. The large availability of datasets and the advancement of technologies like \textit{\acrfull{ml}} and \textit{\acrfull{ai}} to process and analyse such datasets are the reasons for this growth.
The challenge is to develop new types of data analytics, as well as different storage and analysis methods, to handle the volume, velocity, variety, veracity and value of Big Data coming from different sources. Big Data and new horizons of computational powers enable powerful tools like \textit{\acrshort{ml}} and \textit{\acrshort{ai}} to be a part of the everyday-work stack of any data scientist. Those characteristics make it affordable to train bigger and more advanced and even revolutionary \textit{\acrshort{ml}} models. One outstanding result is \textit{GPT3}, a language model with 175 billion parameters (which is 10x more than any previous language model) trained by OpenAI on about 45 terabytes of textual data from different kinds of datasets. This model is capable of understanding human speech in a very deep way, achieving never-seen results in all kinds of language oriented tasks: generating news articles, translation, question answering, and many more \cite{brown2020language}. It is now well established that bigger models, in terms of trained parameters, can generalize better than smaller one, but of course, those require an even larger amount of data \cite{pmlr-v97-brutzkus19b}.

It is, therefore, necessary to make the process of gathering and linking data from many different sources affordable to perform the desired analysis. A common situation for most companies is the availability of their own data collected through their own business activities and the need of integrating more data from external data sources to gather insights.

\begin{figure}
  \centering
  \includegraphics[scale=0.3]{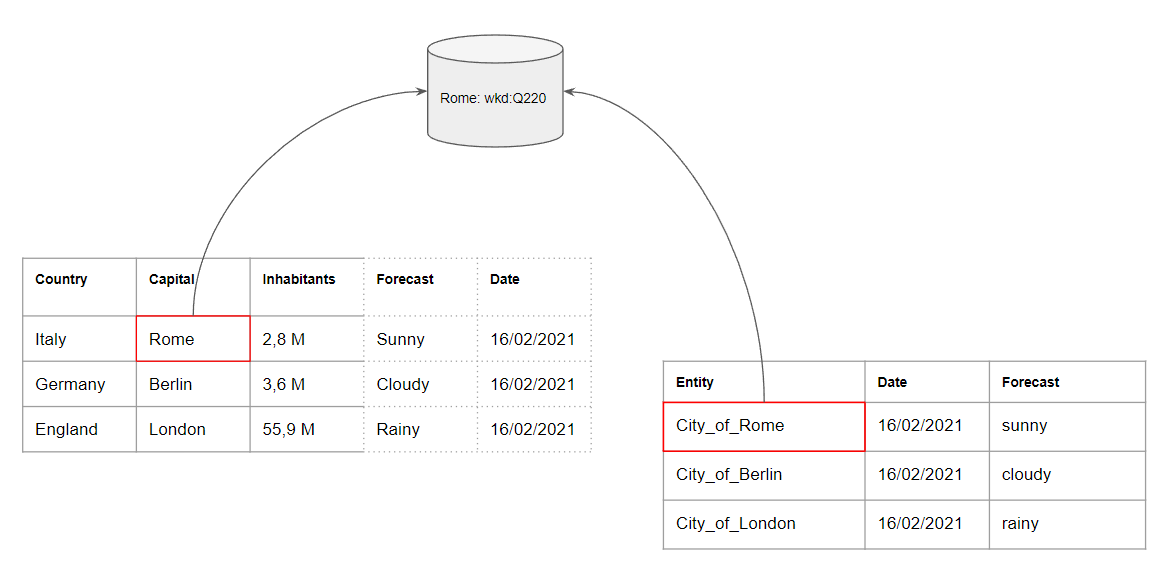}
  \caption{A data enrichment example: weather information added to a capital table.}
  \label{fig:enrichmentExample}
\end{figure}

Data enrichment refers to the process of enhancing existing information by supplementing missing or incomplete data with data from external data sources \cite{azad2018business}. For example, figure \ref{fig:enrichmentExample} describes a situation where a source dataset containing information about cities and touristic attractions could be enriched with weather forecasts to plan future indoor/outdoor activities. A key issue of data enrichment is that data are often ambiguous and it is important to understand how the same entity is identified across data sources, so that the correct information can then be added to the existing set of data. For example,\emph{``Rome''} and \emph{``City\_of\_Rome''} might refer to different entities, e.g., the geographical location or the administrative division.

Semantics assumes an essential role in the process of entity disambiguation across data sources, matching them to a target \acrfull{kg} (e.g., Wikidata) with the aim to associate unique identifiers with entities to distinguish them from one another. While many different automatic algorithms exist to perform the reconciliation task, there are many cases where automation is not enough and human knowledge is required to have the final verdict on ambiguous results. Once the entities have been reconciled, the datasets of interest can be integrated. Continuing the above example, the activity consists of finding a dataset that provides weather information for the cities of interest, understanding the unique identifiers that are used to identify the locations, reconciling the source dataset to get the same identifiers, and finally extending the source dataset by adding the desired weather information.

Despite having countless of powerful tools which help with data preprocessing and preparation, the tasks just described are difficult and rather costly in terms of both time and money, but are essential for any data analysis pipeline. Moreover, those tools do not always target less experienced users but are limited to domain experts.

This introduction presents the motivations that led to building a framework which (i) guides users through the tasks of semantic enrichment of tabular data, (ii) supports the choices made by reconciliation algorithms, with the ability to refine and perfect matches, (iii) and finally offers ways to integrate new external services for both reconciliation and extension tasks.

\subsection{Semantic Table Enrichment}
\emph{Data enrichment} is a particular data integration task which objective is to manipulate a source dataset so that new data from external sources can be integrated in it. Data enrichment is not exactly a data integration task for these main reasons: (i) only the source datasets are known and under the control of the integrator, (ii) there is no control over external data sources, (iii) normally data from different sources are usually heterogeneous. For these reasons, techniques and well-established approaches for data integration, e.g., \emph{Record Linkage}, cannot be adopted to solve the problems of enrichment.

However, semantic data can be exploited to discover links between entities across different datasets and when working on tables, different types of semantic annotations can be associated with them. According to the ``Semantic Web Challenge on Tabular Data to Knowledge Graph Matching'' (SemTab) \cite{jimenez2020semtab}, three annotation tasks can be identified to annotate a table with semantic meaning. \acrfull{cea} is a cell-based annotation task, also known as instance-level annotation. It's the task of annotating the textual values in the table cells (mentions) and it consists of an entity linking algorithm whose objective is to disambiguate mentions from the entities inside a \Acrshort{kg}. \acrfull{cta} is a schema-level annotation task which aims to map the underlying table schema to a \Acrshort{kg} ontology. Finally, \acrfull{cpa} is a schema-level annotation that defines binary relations (properties of the \acrshort{kg} ontology) between pairs of columns of the table.

The ultimate goal of \emph{Semantic Enrichment of Tabular Data} is to find a shared space of identifiers between two or more datasets, so that information can be linked together and used to enrich the initial set of data. Semantic Enrichment introduces two core tasks in the process of enrichment: reconciliation and extension. Reconciliation aims to map the mentions of a table to a target \Acrshort{kg} to set unique identifiers. The task is performed with automatic algorithms that can produce ambiguous results given the ambiguity of table mentions. Human knowledge and judgment are required to address those situations. Once mentions have been mapped to entities of a \Acrshort{kg}, the extension task aims to use unique identifiers to fetch data from different data sources, and add those data to the initial table with an approach based on Linked Data principles \cite{bizer2009emerging}. 

\section{Related works}\label{sota}
Data enrichment mainly consists of a transformation pipeline applied to the involved datasets to integrate data between each other. Many tools exist and are available to expert users in building data transformation pipelines: command line interface tools (e.g., csvki, a powerful command line interface tool for data transformation on tabular data for Python), libraries for data analysis (e.g., Agate a data analysis library optimized for humans), and many desktop and web applications (e.g., Microsoft Excel, Google Sheets) can view and manipulate tabular data. The tools just presented offer many ways to transform and create data transformation pipelines through configurable components, but they do not hint to any kind of integration or enrichment functionality through the use of semantics between data sources.

OpenRefine\footnote{https://openrefine.org/} is a powerful tool for working with messy data through interactive user interfaces available as a desktop application. It provides functionalities for cleaning and transforming data from one format to another, while also offering semantic enrichment processes through reconciliation services offered as Web APIs. OpenRefine is able to connect to any reconciliation service which follows a specific W3C specification \footnote{\url{https://reconciliation-api.github.io/specs/latest/}}. At the moment of writing many services are available: Wikidata, OpenCorporate which contains 171 million corporate entities, Nomisma and Ordnance Survey which provides endpoints to various reference datasets. A full list of currently available services is specified in their GitHub documentation\footnote{https://github.com/OpenRefine/OpenRefine/wiki/Reconcilable-Data-Sources}, while the community still works to constantly add new ones. Although the designed architecture is well documented and supported, the tool still remains accessible to experts in the domain.

Datagraft \cite{roman2018datagraft} is a cloud-based service which provides hosting and transformation of data. It allows users to interactively design data transformation pipelines for tabular data, while also sharing those processes across multiple datasets so that, already defined transformation rules can be applied when the same kind of data is encountered. It allows users to query \Acrshort{kg}s data through SPARQL endpoints or with API services and operate semantic enrichment tasks. 
Datagraft relies on two different tools: Grafterizer \cite{maurino2019modelling} and ASIA \cite{cutrona2019asia}. The first one offers functionality for data preparation, transformation, and cleaning, while the second one provides schema and instance-level annotations through reconciliation services and data extension with extension services. The UI is built with three main components: (i) Tabular transformation provides a pipeline which can be defined and exported to be applied to the same data when encountered again in future tables; (ii) Tabular annotation provides an annotation process that associates the concepts and datatype of a target \Acrshort{kg} with table columns, and reconcile table cells to \Acrshort{kg} entities using ASIA exposed services. Even though the annotation process is manual, the tool offers suggestions about concepts, datatypes and properties using an external service called ABSTAT \cite{spahiu2016abstat}; (iii) RDF mapping enables a manual or automatic process to generate RDF triples from annotations.

Karma \cite{gupta2012karma} is an open source project for data integration. It provides users with data cleaning, data normalization, and data annotation and integration functionalities, guided by a data model based on a vocabulary defined as an ontology. Karma utilizes a database containing annotations obtained during previous executions of the tool to automatically generate a model to apply to future data, allowing users to also refine its behaviour and correct its mistakes.
The annotation process of Karma is mainly composed of two phases: (i) specify semantic types by defining a relation between a table column and a target ontology, also suggesting annotations automatically if a previous similar column has already been annotated, and (ii) define a relation between each annotated columns. The final output of the two-step process can be exported in various formats like RDF and JSON-LD.

Odalic \cite{knap2017towards} is a tool for \Acrshort{sti} based on the TableMiner+ \cite{zhang2017effective} algorithm. Odalic has two inputs: a table of data described in a CSV file and one or more Knowledge Bases that can be combined to improve the results of the annotation.
Odalic enables users to start a task of semantic annotation on the loaded table in a completely automatic manner, unlike previously presented tools. The annotation process provides, as final results, \Acrshort{cea}, \Acrshort{cta} and \Acrshort{cpa}. The UI allows users to make changes to the product of the annotation task to improve the final result. If changes are made to the annotation results, the process is restarted by maintaining constraint imposed by the users.
Finally, as made available from other tools, it is also possible to export the achieved annotation on the table data, in different kinds of formats like RDF, JSON-LD, and CSV with added columns corresponding to the annotation data.

STAN \cite{stan} is a tool available as a web application to annotate tabular data with semantic information. STAN allows users to upload a table in CSV format with the presence of the header and information on the used separator. Once a table is loaded, the task of annotation can be done semi-automatically. Like Datagraft most of the annotation must be given manually, but suggestions are available to the user through the use of ABSTAT. A peculiarity of the tool is that it allows users to define their own ontology during the annotation process, instead of using a known knowledge base. 

Table Miner+ \cite{mazumdar2016tool} is a tool for the \Acrshort{sti} based on its homonym approach \cite{zhang2017effective}. Odalic, also makes use of the Table Miner+ \Acrshort{sti} algorithm which leads to a fully automated annotation approach.
The interface requires the URL of a web page that contains one or more web tables in input, and shows a preview for each table. Through the UI, it is possible to launch an annotation task for one of the identified tables. Because the annotation task may require a long time to complete, the tool also offers the possibility to send a notification e-mail on the completion of the annotation process. 
Results are formatted in JSON and then parsed to be displayed in the two main components that compose the UI. The first component shows the loaded and annotated tables and allows users to interactively explore data and annotation metadata. Each annotation metadata has a confidence score resulted from the annotation process. Concepts (for the columns) and entities (for the table cells) with the highest score are automatically picked by the system, even though the user can select different ones based on the ones retrieved by the automatic process. It is also possible to manually insert a \Acrshort{kg} resource by providing its \Acrshort{uri}.
The second component visualizes annotation  data as a graph. Each row has a contextual action which enable the view of its data in graph format. For example, the header row is shown with as many nodes as many column cells of the table, plus additional nodes for each candidate concept. Links between column nodes represent relations between table columns, while links between node concepts and column cells represent their candidates. It is possible to manipulate links by directly change the matching candidate indicated by a different visual style.

State-of-the-art tools come from the latest SemTab challenge which, starting in 2021, not only assesses STI approaches, but also evaluates the user interface used to perform the annotation tasks. The most notable tools from the challenge are next presented.

DAGOBAH \cite{liu2019dagobah} is a system that performs automatic preprocessing and semantic interpretation of tables. In the latest version, some major improvements are made available in the indexing and matching strategies for entities of the table: it provides a better representation of the entities by exploiting their context in the \Acrshort{kg} which leads to better disambiguation for ambiguous matches. The system is made available through an annotation API and a frontend user interface. Focusing on the UI, it allows users to load new tables into their project and start the preprocessing and annotation pipelines using the \emph{TableAnnotation API}, and finally visualize the results. The system does not allow users to intervene in the reconciliation steps, meaning users are not able to improve the final result.

MAGIC \cite{ongenae2021magic} is presented as a framework capable of both annotating and augmenting a dataset by using a particular embedding technique called Instance Neighboring using Knowledge (INK)\cite{steenwinckel2022ink}. INK is a binary data structure representing a \Acrshort{kg}, taking into account nodes and their neighbourhoods until a certain specified depth K. Based on the interpretable nodes embedding, MAGIC is able to perform the tasks of \Acrshort{cea}, \Acrshort{cta} and \Acrshort{cpa} all at the same time, by also providing additional linked data. The framework is based on a pipeline of steps which leads to the augmentation of semantics for the given table. Additionally, information provided by the INK embeddings of new relations withing the \Acrshort{kg} can be used to enrich the table with information from the same dataset. This system also provides a graphical user interface to augment selected columns. An advantage of this GUI is that it is accessible to non-expert users given its simplicity. While the GUI focuses on the augmentation task, it does not provide any feedback about the annotations retrieved by the process for the tables.

MTab \cite{nguyen2021demonstration} is an automatic tool for tabular data annotation with \Acrshort{kg}s. MTab augments the original table data with schema and instance-level annotations. The tool is able to support multilingual tables and it is able to handle different formats: Excel, CSV, markdown tables. The system presents itself as a series of steps: from preprocessing the tabular data, to enriching the table with semantic annotations with prediction and search tasks. Furthermore, the system also got the 1st place on the SemTab challenge on usability track with a UI that supports the annotation task, showing the results of the automatic process. The UI provides a table upload and an annotate button to start the process composed of the mentioned tasks. It also allows the user to search entities in the most common \Acrshort{kg}s (Wikidata, DBPedia), but it is a standalone feature and it does not help users in any part of the process. 

MantisTable \cite{Cremaschi2020,avogadro2021mantistable} is an unsupervised automatic approach for the Semantic Table Interpretation Task performed against DBpedia and Wikidata. It performs the annotation through the use of LamAPI, a service which allows clients to efficiently fetch data from the Knowledge Graph dumps. An earlier version of SemTUI, called tUI, is proposed as an instrument to interact with the annotation process. The tool is able to load tables in structures called datasets and initiate annotation processes for a specific table or the whole dataset. The UI shows pending processes and some statistics about the tables uploaded to the server. tUI also allows the users to interact with the results obtained from the \Acrshort{sti} process, for example, reviewing candidate entities matched with cell mentions. While it is more complex to use than MTab, it also provides more user interactions aimed to improve the final outcome of the process.

Other approaches were presented at the SemTab 2021, but none of them included a graphical user interface to interact with their novel proposed approaches of semantic table interpretation. Those are simply reported because this work does not focus on developing a \Acrshort{sti} algorithm: GBMTab \cite{yang2021gbmtab}, JenTab \cite{abdelmageed2021jentab} and Kepler-aSI \cite{baazouzi2021kepler}. 

\subsection{Tools Comparison}
Most of the analyzed tools do not include all of the tasks required for the semantic enrichment of tabular data. For example, almost all approaches, DAGOBAH, MTab and MantisTable, presented at the SemTab 2021 do not provide a mean to enrich a table with data from external data sources through the use of semantic annotations (extension task). MAGIC offers a type of augmentation through the use of the INK \Acrshort{kg} representation, even though the extended data comes from the same \Acrshort{kg} represented by the INK embeddings, meaning that external datasets are not involved in the process.
MAGIC UI provides an easy-to-use interface that does not require any particular knowledge about semantics and it is accessible to non-expert users.
MTab UI, the winning contestant for the usability track of the challenge, gives the accessibility advantage of loading tables in various formats and displaying table annotations in a clear way. The tool does not provide any indication about the progress of the annotation process, which could take a long time to complete. It also provides interfaces to search entities contained in common Knowledge Graphs, but they are not employed in supporting and facilitating the user to correct mistakenly assigned annotations.

Only some of the analyzed tools offer the ability to refine matches from the results obtained after an automatic annotation process. As already outlined, it is essential to engage users in the loop of the enrichment process to improve the final outcome. Table Miner+, Odalic and MantisTable provide ways to view, correct and perfect annotations, improving the overall result.
Both Datagraft and STAN, even though they do not provide fully automated approaches for the annotation of a table, offer concepts and relations suggestions to the user through ABSTAT, an external service provided with web APIs.
OpenRefine does not offer support for  \Acrshort{cpa} and structural annotations, i.e.: identifying the subject and objects of relations, meanwhile all other tools allow users to manually or automatically augment the table data with those information. Finally, OpenRefine can integrate new reconciliation services built following the W3C standard specification.
For data preparation, Karma, OpenRefine and Datagraft provide functionalities to design transformation pipelines to apply to the table data before augmenting it with semantic notions.
In particular, Datagraft supports the definition of transformation rules that can be exported and augmented to annotate future tables with the same kind of data structure.

\section{Approach and Implementation}
From the analysis conducted in section \ref{sota}, we can conclude that the development of a comprehensive framework can be relevant to make the enrichment process effective. In this chapter are presented the requirements and architecture of the implemented framework, which is composed of two major components: a user interface to support user interaction, and a backend to manage tables and interface the external annotation services.

\subsection{Requirements Analysis}
The previous analysis of related and state-of-the-art tools revealed requirements that would have to be met by the implementation of the framework in order to fully encapsulate the entire semantic enrichment experience:
\begin{itemize}
    \item It should include the core tasks of the semantic enrichment including both reconciliation and extension.
    \item It should engage the user in the loop of the enrichment process through interactive choices and visualization, so that results from automatic algorithms can be better monitored and, if necessary, corrected.
    \item It should not depend on any specific service, instead, external services, e.g., annotation and extension services, should be easily added to form an ecosystem of services available to end users.
\end{itemize}
The detailed requirements of the framework  are presented in the following sections.

\subsubsection{User Requirements}
User requirements are product features or functions that developers must implement to enable users to accomplish their tasks. In the following list, the identified user requirements are described:
\begin{itemize}
    \item The user must be able to manage tables though basic operations: upload and delete one or more tables in different formats (e.g., CSV, JSON and JSON following the W3C representation for an annotated table), search tables by name, collect information on tables (e.g., number of rows and columns, last modification date), save and reload tables to resume working sessions, and export tables in different formats (e.g., W3C annotation standard format).
    \item The system must provide users with the operations of selecting row, column, and table cells and apply operations on selected entities. Basic (e.g., row and column deletion, cell renaming, and undo/redo of the last operation), filtering (e.g., show only rows with uncertain reconciled cells), and searching (e.g., search for specific cell labels or annotation metadata) operations must be supported.
    \item The reconciliation and extension tasks must be supported by multiple services that a user can select and invoke through the user interface.
    \item The user interface must reflect the status of the annotations. For example, when a cell has been reconciled, a green badge is placed next to the cell content, otherwise a red badge is shown. 
    \item Each table element must be inspectable to show the details of the annotations associated with it. For example, a cell can be inspected to show all annotation candidate entities associated with it, along with details for each of them (e.g., \acrshort{kg}'s id, name, description and score, types and properties). Relations (i.e., properties) between columns can be inspected in a graph-like view by showing relation edges on top of the table view.
    \item The annotation services might return a set of candidate annotations and possibly select one of them as the correct one. The system must allow the user to change the selection, or manually add a new correct entity in the list of candidates.
\end{itemize}

\subsubsection{System Requirements}
System requirements are the constraints imposed on the system. They deal with issues like maintainability and performance. In the following list are described the key system requirements:

\begin{itemize}
    \item The system should be able to handle a large number of elements, which means datasets composed of many tables, tables composed of many rows and/or columns, and large sets of annotation candidates. The user interface must let the user visualize and navigate such elements, and the backend must be able to store annotated datasets and tables.
    \item According to the agile development approach, both the user interface and the backend must be modular to support the introduction of new functionalities, revision or replacement of existing ones by simply changing or replacing components.  
    \item The system must be able to interact with external services and exchange annotation data, therefore the interchange data format should be compliant with standards. The adopted standards are JSON for the representation of tables and the W3C Reconciliation format for annotations\footnote{\url{https://reconciliation-api.github.io/specs/0.1/}}.
    \item A key feature of the system must be the capability of including current and future transformation/reconciliation/extension services with minimal effort, therefore the backend must foresee in and out transformation components that can be specialized to interact with external services. The inclusion of new services must provide immediate feedback to the UI by feeding information to automatically create widgets to interact with the integrated services.
\end{itemize}

\subsection{Framework Architecture}
SemTUI combines the power of a back-end server with the flexibility of a web-based front-end application.
This structure promotes the separation of concerns principle by adding an abstraction layer between the integration of different services and the user interface that allows the access to them.

\begin{figure}[h!]
  \centering
  \includegraphics[scale=0.40]{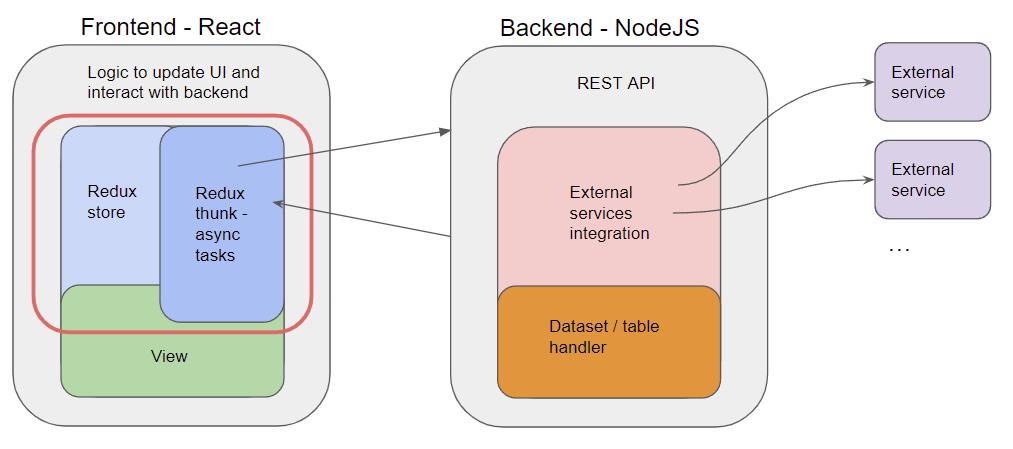}

  \caption{General architecture of SemTUI.}
  \label{fig:architecture}
\end{figure}
The architecture of the framework is shown in figure \ref{fig:architecture}. It is designed to satisfy the presented requirements, and it is divided into two main components:

\begin{itemize}
  \item A frontend web application to handle the interaction with end users and provide the functionalities necessary to the Semantic Enrichment of a table. It is a web application composed of sub-layers to handle the application logic and internal store state, communicate to the backend server through a REST API, and update UI components based on changes to the internal state applied through the user interaction. It is built using React and Redux JavaScript libraries.
  \item A backend server to handle the integration and communication with any of the integrated external (reconciliation and extension) services. New services can be added by writing transformers to get requests from the frontend, query external services, and provide responses in a standard format. The backend also handles the storage and retrieval of tables for the users. All of the functionalities are exposed as a REST API to support the communication with the frontend. It is built using NodeJS JavaScript library and Express JavaScript framework.
\end{itemize}

Figure \ref{fig:workflow}  shows the main operations triggered by the user actions to outline how the framework components interact between each other to satisfy the user requests:

\begin{enumerate}
  \item The user can start the interaction with the system by either selecting one of the uploaded tables, or by uploading a new one. Once a table is selected, the user can perform actions on it. There are two kinds of actions: local actions, e.g., modify the label of a table cell  or visualize its annotation, and actions that send a request to the backend, e.g., save the table, query an annotation service (1).
  \item Local actions simply update the UI application store applying changes to the involved UI components (2.1).
  \item Non-local actions send requests to the backend server (2.2) that queries the external annotation service chosen by the user (3.1), or updates the current table (3.2).
  
\begin{figure}[h!]
  \centering
  \includegraphics[scale=0.45]{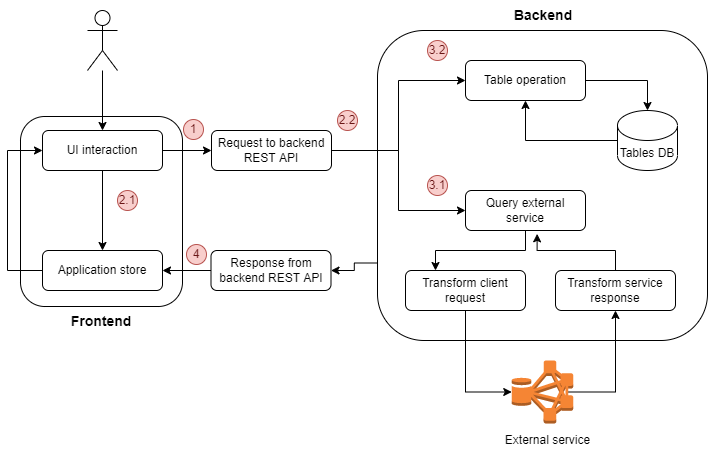}

  \caption{Interactions between the framework components of SemTUI.}
  \label{fig:workflow}
\end{figure}

  \item The backend returns data in the standard format expected by the UI to update the application store (4) that can in turn update the user interface. The data exchanged between the two components is represented in JSON and modeled following the W3C specification for the representation of semantic annotations.
  \end{enumerate}
  
For example, the user may require to query an external reconciliation service which provides an entity annotation for a set of cell labels (instance-level annotation). For this kind of request, a backend pipeline is built to ease the integration of new services. It is composed of an initial transformation of the request from the web app, adapting it to the schema necessary to query the external service. Once a response from the service is obtained, a new transformation is applied to transform it into the standard format expected by the UI.

The UI always receives a standardized response, so it can easily update the application store no matter of the used external service because the response will always be in the same format. Once the application store is updated, the application re-renders its content to reflect the new application state to the UI.

\subsubsection{Reconciliation}
Reconciliation is the key task of the Semantic Enrichment process. In particular, the extension step is highly dependent on the results obtained from the reconciliation, therefore it is important to ensure correctness in the table annotations. SemTUI offers multiple ways to reconcile table mentions through the use of external reconciliation services. 

A reconciled cell presents a marker beside its label identifying the level of annotation. SemTUI proposes the users with five different annotation identifiers to better understand the results obtained from the queried service. If a cell has a candidate matching label, its label is also linked to the \Acrshort{kg} resource. 

\begin{figure}[h!]
  \centering
  \includegraphics[scale=0.6]{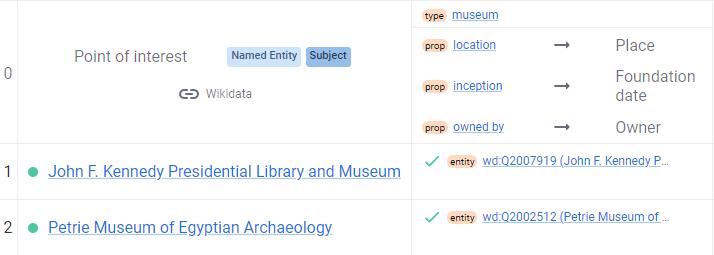}

  \caption{Example of compact annotation visualization for a column and their cells.}
  \label{fig:exampleCompactAnnotation}
\end{figure}

\begin{figure}[h!]
  \centering
  \includegraphics[scale=0.27]{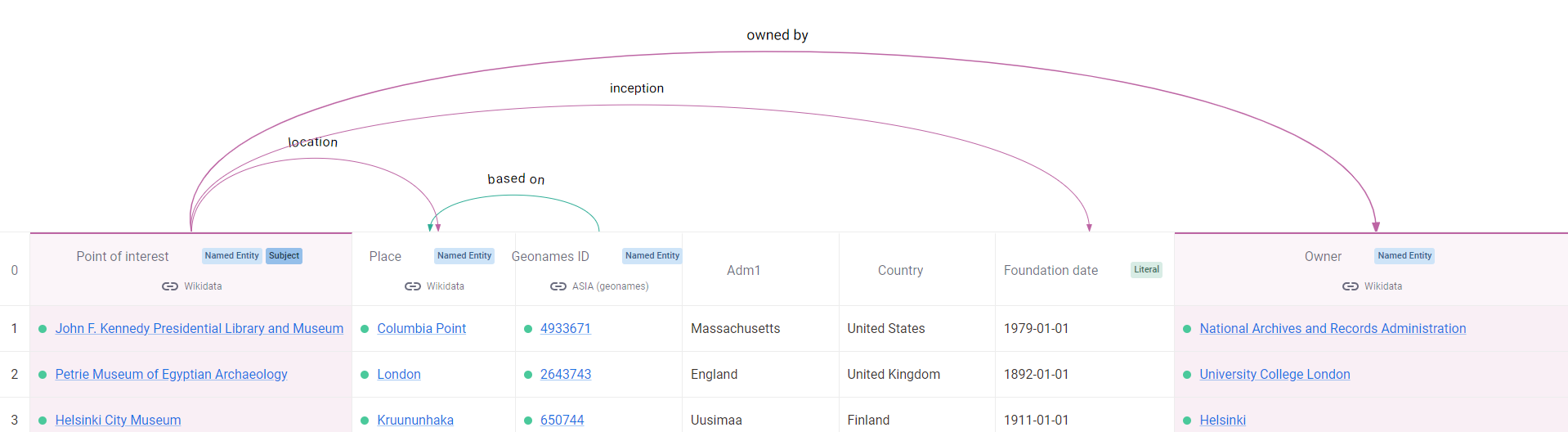}

  \caption{SemTUI visualization of column properties.}
  \label{fig:propertiesVisualization}
\end{figure}

SemTUI offers a compact visualization for the annotation of both the columns and cells. Figure \ref{fig:exampleCompactAnnotation} shows an example where the column \emph{Point of interest} is annotated as the subject column and contains named entities of type \emph{museum}. The column has also properties, i.e., \emph{location}, \emph{inception} and \emph{owned by}, with objects other columns of the table, i.e., \emph{Place}, \emph{Foundation date} and \emph{Owner}, respectively. Each column mention is then associated with a \Acrshort{kg} entity. SemTUI also offers a property visualization feature that helps users to immediately identify the relations between the columns of the table, an example is shown in figure \ref{fig:propertiesVisualization}.

\subsubsection{Annotation Refinement}
SemTUI adds another task to the Semantic Enrichment process of a table, called \emph{Refinement}.
\emph{Refinement} is a user-centred task in which the combination of human knowledge and interactivity helps to solve the ambiguous results produced by automatic annotation algorithms to deliver accurate annotations and guarantee a correct extension with external data.

The first set of refinement features include row and cell filtering based on diverse criteria. A search functionality offers the user a filtering criterion based on the selected filter, e.g., match by annotation metadata name and type. The search is applied globally to the table cells. If a row in the table is matched with the applied filter, the row is kept highlighting the matching cells, otherwise the row will be filtered out of the visualization.

\begin{figure}[h!]
  \centering
  \includegraphics[scale=0.5]{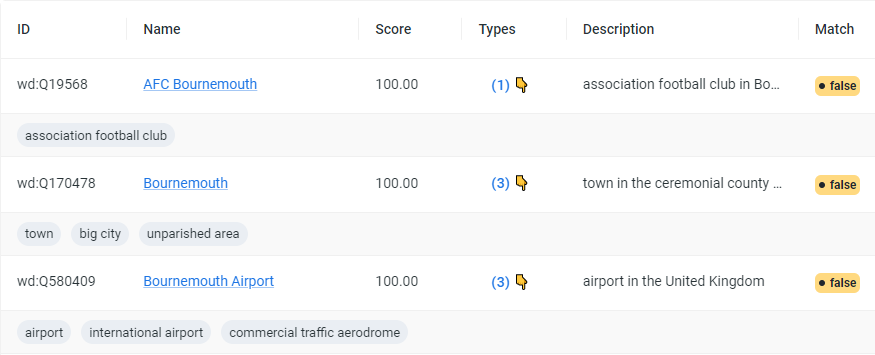}

  \caption{Inspecting a cell candidate entities. The reconciliation service used provides types and descriptions in additions to required data fields (id, name, score, match).}
  \label{fig:exampleInspect}
\end{figure}

The table cells can also be inspected manually by the user, so that a more accurate revision of the associated annotations can be made. Each candidate entity associated with a table mention can be selected by the user, so that a match can be corrected if the automatic algorithm mistakenly assigned a wrong matching entity, or the candidates were so ambiguous to prevent any automatic decision taken by the reconciliation service. Figure \ref{fig:exampleInspect} shows an example of the candidate entities while inspecting a cell. In the example, the cell contains the label \emph{Bournemouth}. The \emph{Wikidata-OpenRefine} service that was used returned multiple candidates with the same score, so it was not possible to assign a true match to either of them. The user can use his knowledge to notice that the column where the cell is located contains cells that refers to English football clubs. It is therefore clear that the first candidate entity of type ``\emph{association football club ...}'' is the correct match for the cell label.

SemTUI also features processes that offer a more automatic way to refine the candidate matches currently annotated in the table. For example, refinement by entity type, i.e., assign a true candidate match based on the selected entity types, or by entity score, i.e., assign a true candidate match based on a selected score threshold.

\subsubsection{Extension}
Once at least some entities in a column have been reconciled, it is possible to extend the table by adding new columns fed with external data. SemTUI provides, just like the reconciliation task, external services which enable the extension with external datasets. 

While the majority of the currently available tools do not even offer the possibility of the extension, SemTUI makes the task as easier as possible. Just like the semantic annotation of the table, the extension is not strictly related to a particular algorithm or service, instead the framework aggregates multiple services building an extensible infrastructure where the user is free to choose the best service for his needs. Currently, the available services are limited and more research work to support the task is needed.
\section{Efficiency and Usability Testing}\label{usability}
The usability and efficiency of the system has been tested following a standard evaluation process constituted by guided user tests followed by a user experience questionnaire administered to the participants.

In a usability testing session, the evaluator asked a participant to perform a series of tasks involving the functionalities of SemTUI. While the participant worked towards completing the described tasks, the evaluator observed the participant's behavior, keeping track of the time to complete a task, and listening to any feedback.

The testing experiment has been conducted to:
\begin{itemize}
  \item Identify problems in the design of features and components of the product.
  \item Expose and uncover any opportunity to improve the current prototype.
  \item Learn how the users usually behave with the application, enabling developers to adapt future changes, so that the user's behavior is considered.
\end{itemize}
Two types of participants have been gathered to perform the evaluation process and represent a sample of the target population:

\begin{itemize}
  \item Non-expert users: users with no experience in the domains of semantics, knowledge graphs or data integration.
  \item Expert users: users aware of semantics and related topics, usually working with data from multiple datasets to perform analysis tasks.
\end{itemize}
A total of 22 participants were involved in the evaluation, with 11 non-expert users and 11 more experienced users. Participants were considered expert or non-expert users according to what they have declared in the questionnaire about familiarity with semantics, knowledge graphs, and data integration or enrichment.

The experiment has been conducted by deploying a remote version of the framework, to ensure that any request to the external services is executed within the same time frame, this way any external variable can be excluded from the latency of the network from the user's client to the backend server.

\subsection{User Experience Questionnaire}
To assess the user experience we adopted a standard and reliable user experience questionnaire called \emph{UEQ}\footnote{https://www.ueq-online.org/}.
The questionnaire consists of 26 items carefully constructed to measure six different user experience scales. Each item is represented by two terms with opposite meaning and a rating scale between them. The order of the terms is intentionally randomised among the items, so that inconsistencies and suspicious data that users have answered are easier to detect.

A seven-stage scale is used to reduce the central tendency bias, i.e., reduce the tendency for a rater to place a neutral vote in the middle of the scale. The items are scaled from -3 to +3, where -3 is the most negative answer, 0 is a neutral answer, and +3 is the most positive answer.

The UEQ measures the user experience based on six scales resulted from of a study conducted on 11 usability tests involving about 900 participants \cite{laugwitz2008construction}:

\begin{itemize}
  \item Attractiveness: the overall impression of a product.
  \item Perspicuity: how easy is to learn and get familiar with the product?
  \item Efficiency: how much effort is required by the users to complete their tasks?
  \item Dependability: is the user free and in control of his actions?
  \item Stimulation: does product arouse excitement?
  \item Novelty: how innovative and creative is the product?
\end{itemize}
\emph{Attractiveness} has six items and generally provides an answer to the question \emph{"Do users like or dislike the product?"}. Attractiveness is then divided into the pragmatic qualities of the product, i.e: aspects purely goal-directed, and hedonic qualities of the product, i.e.: not related to any goal. Pragmatic qualities include \emph{Perspicuity}, \emph{Efficiency} and \emph{Dependability}. Hedonic qualities include \emph{Stimulation} and \emph{Novelty}.

\subsection{Results}
Results have been gathered from 22 participants, mostly being students and researchers, and few professors. The number of participants has been chosen according to the UEQ specification, so that the scale means of the answered items of the questionnaire can be interpreted with a certain precision. Table \ref{tab:precision} shows the rough estimation of the sample size based on the UEQ benchmark and standard deviation of the gathered samples. In yellow are highlighted the results of this experiment. Each item of the table identifies an estimation of the size of the sample, so that the specific scale can be measured with a specific precision and error probability, \emph{0.5} and \emph{0.01} in our case. 

\begin{table}[htb]
 \centering
  \caption{The result of the evaluation tests.}
 \scalebox{0.65}{
 \begin{tabular}{|l|c|c|c|c|c|c|} 
 \hline
 \rowcolor[rgb]{0.502,0.502,0.502} \textcolor{white}{Condition} & \textcolor{white}{Attractiveness} & \textcolor{white}{Perspicuity} & \textcolor{white}{Efficiency} & \textcolor{white}{Dependability} & \textcolor{white}{Stimulation} & \textcolor{white}{Novelty} \\ 
 \hline
 {\cellcolor[rgb]{0.502,0.502,0.502}}\textcolor{white}{Precision=0.5, Err.Prob.=0.1} & 6 & 9 & 7 & 6  & 7 & 11 \\ 
 \hline
 {\cellcolor[rgb]{0.502,0.502,0.502}}\textcolor{white}{Precision=0.5, Err.Prob.=0.05} & 9 & 13 & 9 & 8  & 10 & 15 \\ 
 \hline
 \rowcolor[rgb]{1,0.992,0.8} {\cellcolor[rgb]{0.655,0.643,0.447}}\textcolor{white}{Precision=0.5, Err.Prob.=0.01} & 16  & 22 & 16  & 14  & 17 & 22 \\ 
 \hline
 {\cellcolor[rgb]{0.502,0.502,0.502}}\textcolor{white}{Precision=0.25, Err.Prob.=0.1} & 25  & 37 & 27  & 23  & 27 & 42 \\ 
 \hline
 {\cellcolor[rgb]{0.502,0.502,0.502}}\textcolor{white}{Precision=0.25, Err.Prob.=0.05}  & 36  & 53 & 38  & 32  & 39 & 60 \\ 
 \hline
 {\cellcolor[rgb]{0.502,0.502,0.502}}\textcolor{white}{Precision=0.25, Err.Prob.=0.01}  & 62  & 91 & 65  & 56  & 67 & 103 \\ 
 \hline
 {\cellcolor[rgb]{0.502,0.502,0.502}}\textcolor{white}{Precision=0.1, Err.Prob.=0.1} & 159  & 233  & 167  & 143 & 171  & 264 \\ 
 \hline
 {\cellcolor[rgb]{0.502,0.502,0.502}}\textcolor{white}{Precision=0.1, Err.Prob.=0.05} & 224  & 329  & 236  & 201 & 241  & 372 \\ 
 \hline
 {\cellcolor[rgb]{0.502,0.502,0.502}}\textcolor{white}{Precision=0.1, Err.Prob.=0.01} & 388  & 569  & 409  & 349 & 418  & 645 \\
 \hline
 \end{tabular}
 }
 \label{tab:precision}
\end{table}

\subsubsection{Efficiency}
\begin{figure} [h]
  \centering
  \includegraphics[scale=0.40]{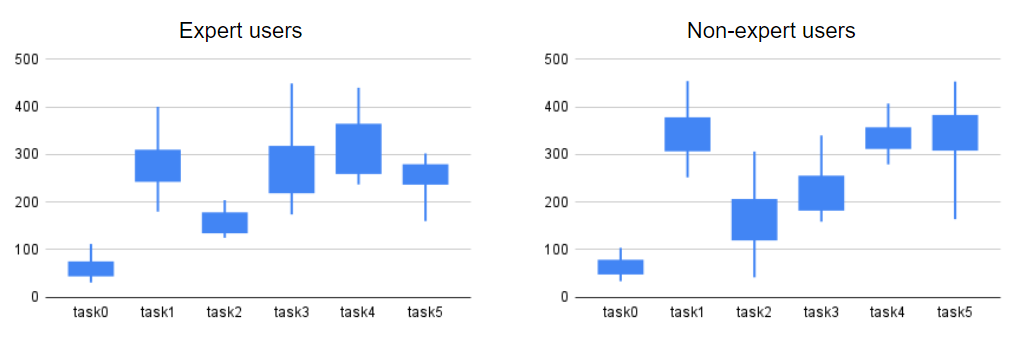}
  \caption{The time in seconds taken to complete the tasks by the participants.}
  \label{fig:timeDivided}
\end{figure}

The first analysis conducted was relative to the application efficiency in completing the user tasks. For each participant the evaluator kept track of the time necessary to complete the requested user task. In table \ref{tab:timeAll} the average completion time (Avg time), the standard deviation time (Std time) and the reference time has been reported for each task. The reference time has been computed as the average time between a slow, normal, and fast execution of each task by the evaluator. They are intended to provide a general idea with respect to the obtained results.
The time difference in completing the tasks between expert and non-expert users using a box plot is represented in figure \ref{fig:timeDivided}.

\begin{table}[h!]
 \centering
  \caption{\label{tab:timeAll}The average completion time required to complete each task.}
 \scalebox{0.65}{
 \begin{tabular}{|r|l|l|l|l|l|l|} 
 \hline
 \multicolumn{1}{|l|}{} & \textbf{task0} & \textbf{task1} & \textbf{task2} & \textbf{task3} & \textbf{task4} & \textbf{task5} \\  
 \hline
\rowcolor[rgb]{0.859,0.859,0.859} \multicolumn{1}{|l|}{\textbf{Avg time}} & 1m4s & 5m12s & 2m40s & 4m15s & 5m25s & 4m54s \\ 
\hline
\rowcolor[rgb]{0.859,0.859,0.859} \multicolumn{1}{|l|}{\textbf{Std time}} & 23s & 1m7s & 56s & 1m13s & 56s & 1m17s \\ 
\hline
\rowcolor[rgb]{0.984,1,0.761} \multicolumn{1}{|l|}{\textbf{Reference time}} & 58s & 3m54s & 2m42s & 3m20s & 4m3s & 3m59s \\
\hline
 \end{tabular}
 }
\end{table}

The time tracking results (table \ref{tab:timeAll}) show that the enrichment tasks (reconciliation, refinement and extension) can be completed in few minutes. 
The same tasks performed without SemTUI would require the invocation of single reconciliation services or even searching the \Acrshort{kg}, the manual extraction of extension data from reconciled datasets, and finally perform the manual integration in a final reconciled and extended table. Such a process would require hours compared to the minutes required with the help of SemTUI. 

Additionally, 6 participants, among those with times worse than the reference times shown in the table \ref{tab:timeAll}, were selected to repeat the experiment to better evaluate how easy the tool is to learn. Both expert and non-expert users are involved in second test. Table \ref{tab:timeExperiments} shows the actual time to complete the tasks in the first and second experiment. The difference between the two categories of users is negligible, showing that the use of the tool can be learned quickly.
Finally, about 70\% (15/22) of participants, including 50\% of the non-expert users, showed enough confidence to repeat all tasks without using the tutorial feature.

\begin{table}
\centering
\caption{\label{tab:timeExperiments}The time in seconds taken to complete the tasks in the first (left) and second (right) experiment.}
\scalebox{0.65}{
\begin{tabular}{|r|l|l|l|l|l|l|l|ll|r|l|l|l|l|l|l|l|} 
\cline{1-8}\cline{11-18}
\multicolumn{1}{|l|}{\textbf{user id}}                                      & \textbf{user type}                   & \textbf{task0}                            & \textbf{task1}                            & \textbf{task2}                            & \textbf{task3}                            & \textbf{task4}                            & \textbf{task5}                            &  &  & \multicolumn{1}{l|}{\textbf{\textbf{user id}}}                                      & \textbf{\textbf{user type}}          & \textbf{\textbf{task0}}                & \textbf{\textbf{task1}}                   & \textbf{\textbf{task2}}                   & \textbf{\textbf{task3}}                   & \textbf{\textbf{task4}}                   & \textbf{\textbf{task5}}                    \\ 
\cline{1-8}\cline{11-18}
4                                                                           & E                                    & 1m52s                                     & 4m8s                                      & 3m1s                                      & 6m2s                                      & 4m12s                                     & 4m35s                                     &  &  & 4                                                                                   & E                                    & 1m2s                                   & 3m40s                                     & 2m48s                                     & 3m15s                                     & 4m8s                                      & 4m22s                                      \\ 
\cline{1-8}\cline{11-18}
6                                                                           & E                                    & 1m24s                                     & 3m57s                                     & 2m5s                                      & 5m6s                                      & 4m26s                                     & 4m3s                                      &  &  & 6                                                                                   & E                                    & 1m10s                                  & 3m10s                                     & 2m10s                                     & 3m12s                                     & 4m6s                                      & 4m9s                                       \\ 
\cline{1-8}\cline{11-18}
11                                                                          & E                                    & 40s                                       & 3m                                        & 3m24s                                     & 5m13s                                     & 7m                                        & 2m40s                                     &  &  & 11                                                                                  & E                                    & 45s                                    & 2m58s                                     & 3m2s                                      & 3m30s                                     & 4m23s                                     & 2m56s                                      \\ 
\cline{1-8}\cline{11-18}
13                                                                          & NE                                   & 1m2s                                      & 6m59s                                     & 3m50s                                     & 4m2s                                      & 5m21s                                     & 6m2s                                      &  &  & 13                                                                                  & NE                                   & 1m5s                                   & 4m22s                                     & 3m5s                                      & 3m33s                                     & 5m2s                                      & 5m15s                                      \\ 
\cline{1-8}\cline{11-18}
19                                                                          & NE                                   & 1m44s                                     & 5m15s                                     & 2m5s                                      & 3m2s                                      & 5m24s                                     & 5m                                        &  &  & 19                                                                                  & NE                                   & 59s                                    & 4m10s                                     & 2m13s                                     & 3m12s                                     & 5m11s                                     & 4m23s                                      \\ 
\cline{1-8}\cline{11-18}
20                                                                          & NE                                   & 1m40s                                     & 5m6s                                      & 2m46s                                     & 3m11s                                     & 6m47s                                     & 7m33s                                     &  &  & 20                                                                                  & NE                                   & 1m3s                                   & 5m12s                                     & 2m44s                                     & 3m27s                                     & 4m58s                                     & 4m49s                                      \\ 
\hhline{|--------~~--------|}
\multicolumn{1}{|l|}{{\cellcolor[rgb]{0.51,0.51,0.51}}}                     & {\cellcolor[rgb]{0.51,0.51,0.51}}    & {\cellcolor[rgb]{0.51,0.51,0.51}}         & {\cellcolor[rgb]{0.51,0.51,0.51}}         & {\cellcolor[rgb]{0.51,0.51,0.51}}         & {\cellcolor[rgb]{0.51,0.51,0.51}}         & {\cellcolor[rgb]{0.51,0.51,0.51}}         & {\cellcolor[rgb]{0.51,0.51,0.51}}         &  &  & \multicolumn{1}{l|}{{\cellcolor[rgb]{0.51,0.51,0.51}}}                              & {\cellcolor[rgb]{0.51,0.51,0.51}}    & {\cellcolor[rgb]{0.51,0.51,0.51}}      & {\cellcolor[rgb]{0.51,0.51,0.51}}         & {\cellcolor[rgb]{0.51,0.51,0.51}}         & {\cellcolor[rgb]{0.51,0.51,0.51}}         & {\cellcolor[rgb]{0.51,0.51,0.51}}         & {\cellcolor[rgb]{0.51,0.51,0.51}}          \\ 
\hhline{|--------~~--------|}
\multicolumn{1}{|l|}{{\cellcolor[rgb]{0.859,0.859,0.859}}\textbf{Avg time}} & {\cellcolor[rgb]{0.859,0.859,0.859}} & {\cellcolor[rgb]{0.859,0.859,0.859}}1m23s & {\cellcolor[rgb]{0.859,0.859,0.859}}4m44s & {\cellcolor[rgb]{0.859,0.859,0.859}}2m51s & {\cellcolor[rgb]{0.859,0.859,0.859}}4m26s & {\cellcolor[rgb]{0.859,0.859,0.859}}5m31s & {\cellcolor[rgb]{0.859,0.859,0.859}}4m58s &  &  & \multicolumn{1}{l|}{{\cellcolor[rgb]{0.859,0.859,0.859}}\textbf{\textbf{Avg time}}} & {\cellcolor[rgb]{0.859,0.859,0.859}} & {\cellcolor[rgb]{0.859,0.859,0.859}}1m & {\cellcolor[rgb]{0.859,0.859,0.859}}3m55s & {\cellcolor[rgb]{0.859,0.859,0.859}}2m40s & {\cellcolor[rgb]{0.859,0.859,0.859}}3m21s & {\cellcolor[rgb]{0.859,0.859,0.859}}4m38s & {\cellcolor[rgb]{0.859,0.859,0.859}}4m19s  \\ 
\hhline{|--------~~--------|}
\multicolumn{1}{|l|}{{\cellcolor[rgb]{0.859,0.859,0.859}}\textbf{Std time}} & {\cellcolor[rgb]{0.859,0.859,0.859}} & {\cellcolor[rgb]{0.859,0.859,0.859}}27s   & {\cellcolor[rgb]{0.859,0.859,0.859}}1m22s & {\cellcolor[rgb]{0.859,0.859,0.859}}42s   & {\cellcolor[rgb]{0.859,0.859,0.859}}1m12s & {\cellcolor[rgb]{0.859,0.859,0.859}}1m9s  & {\cellcolor[rgb]{0.859,0.859,0.859}}1m40s &  &  & \multicolumn{1}{l|}{{\cellcolor[rgb]{0.859,0.859,0.859}}\textbf{\textbf{Std time}}} & {\cellcolor[rgb]{0.859,0.859,0.859}} & {\cellcolor[rgb]{0.859,0.859,0.859}}8s & {\cellcolor[rgb]{0.859,0.859,0.859}}49s   & {\cellcolor[rgb]{0.859,0.859,0.859}}23s   & {\cellcolor[rgb]{0.859,0.859,0.859}}9s    & {\cellcolor[rgb]{0.859,0.859,0.859}}29s   & {\cellcolor[rgb]{0.859,0.859,0.859}}47s    \\
\hhline{|--------~~--------|}
\end{tabular}
}
\end{table}

\subsubsection{Usability}
The UEQ questionnaire measures the usability and user experience of a tool using six different scales, as described above. For each item of the questionnaire and for each scale, means have been computed with 95\% confidence intervals. 

Table \ref{tab:singleItems} shows the results for each item of the questionnaire. Following the UEQ specification, items with a mean value between -0.8 and 0.8 represent a more or less neutral evaluation with respect to their corresponding scale, values above 0.8 represent a positive evaluation and values below -0.8 represent a negative evaluation. The range of the scales is between -3 (horribly bad) and +3 (excellent).
The results show that 25/26 items received a positive evaluation. The only item with a neutral evaluation is the item \emph{13} corresponding to \emph{complicated / easy} belonging to the \emph{Perspicuity} scale.
The items can then be grouped in the six user experience scales already described
Table \ref{tab:scaleItems} shows the means, standard deviations, and confidence intervals for each scale.

\begin{table}[h!]
\centering
\caption{\label{tab:singleItems}Mean for each item in the questionnaire. Green means positive evaluation, and yellow neutral.}
\label{tab:singleItems}
\scalebox{0.60}{
\begin{tabular}{|c|c|c|c|l|l|l|c|} 
\hline
\multicolumn{8}{|c|}{{\cellcolor[rgb]{0.502,0.502,0.502}}\textcolor{white}{Confidence interval (p=0.05) per item }}                                                                                                                                                                                                                                             \\ 
\hline
\rowcolor[rgb]{0.502,0.502,0.502} \textcolor{white}{Item} & \multicolumn{1}{l|}{\textcolor{white}{Mean}} & \multicolumn{1}{l|}{\textcolor{white}{Variance}} & \multicolumn{1}{l|}{\textcolor{white}{Std. Dev.}} & \textcolor{white}{Confidence} & \multicolumn{2}{l|}{\textcolor{white}{Confidence interval }} & \multicolumn{1}{l|}{\textcolor{white}{Scale}}  \\ 
\hline
1                                                         & {\cellcolor[rgb]{0.741,1,0.792}}1,5          & 0,8                                              & 0,9                                               & 0,381                         & 1,074  & 1,836                                               & Attractiveness                                 \\ 
\hline
2                                                         & {\cellcolor[rgb]{0.741,1,0.792}}1,1          & 1,4                                              & 1,2                                               & 0,488                         & 0,649  & 1,624                                               & Perspicuity                                    \\ 
\hline
3                                                         & {\cellcolor[rgb]{0.741,1,0.792}}1,2          & 2,1                                              & 1,4                                               & 0,604                         & 0,623  & 1,831                                               & Novelty                                        \\ 
\hline
4                                                         & {\cellcolor[rgb]{0.741,1,0.792}}1,1          & 2,1                                              & 1,4                                               & 0,604                         & 0,487  & 1,695                                               & Perspicuity                                    \\ 
\hline
5                                                         & {\cellcolor[rgb]{0.741,1,0.792}}1,7          & 1,4                                              & 1,2                                               & 0,486                         & 1,242  & 2,213                                               & Stimulation                                    \\ 
\hline
6                                                         & {\cellcolor[rgb]{0.741,1,0.792}}1,0          & 1,6                                              & 1,3                                               & 0,532                         & 0,468  & 1,532                                               & Stimulation                                    \\ 
\hline
7                                                         & {\cellcolor[rgb]{0.741,1,0.792}}2,1          & 0,8                                              & 0,9                                               & 0,371                         & 1,765  & 2,508                                               & Stimulation                                    \\ 
\hline
8                                                         & {\cellcolor[rgb]{0.741,1,0.792}}1,5          & 1,4                                              & 1,2                                               & 0,495                         & 0,960  & 1,949                                               & Dependability                                  \\ 
\hline
9                                                         & {\cellcolor[rgb]{0.741,1,0.792}}1,8          & 1,7                                              & 1,3                                               & 0,546                         & 1,227  & 2,319                                               & Efficiency                                     \\ 
\hline
10                                                        & {\cellcolor[rgb]{0.741,1,0.792}}1,2          & 2,0                                              & 1,4                                               & 0,590                         & 0,637  & 1,817                                               & Novelty                                        \\ 
\hline
11                                                        & {\cellcolor[rgb]{0.741,1,0.792}}1,9          & 1,4                                              & 1,2                                               & 0,498                         & 1,411  & 2,407                                               & Dependability                                  \\ 
\hline
12                                                        & {\cellcolor[rgb]{0.741,1,0.792}}2,2          & 0,5                                              & 0,7                                               & 0,286                         & 1,941  & 2,514                                               & Attractiveness                                 \\ 
\hline
13                                                        & {\cellcolor[rgb]{0.969,1,0.741}}0,1          & 2,0                                              & 1,4                                               & 0,595                         & -0,459 & 0,731                                               & Perspicuity                                    \\ 
\hline
14                                                        & {\cellcolor[rgb]{0.741,1,0.792}}1,3          & 1,6                                              & 1,2                                               & 0,522                         & 0,796  & 1,840                                               & Attractiveness                                 \\ 
\hline
15                                                        & {\cellcolor[rgb]{0.741,1,0.792}}0,9          & 1,6                                              & 1,2                                               & 0,521                         & 0,343  & 1,384                                               & Novelty                                        \\ 
\hline
16                                                        & {\cellcolor[rgb]{0.741,1,0.792}}1,6          & 0,9                                              & 1,0                                               & 0,398                         & 1,238  & 2,035                                               & Attractiveness                                 \\ 
\hline
17                                                        & {\cellcolor[rgb]{0.741,1,0.792}}1,5          & 1,4                                              & 1,2                                               & 0,495                         & 0,960  & 1,949                                               & Dependability                                  \\ 
\hline
18                                                        & {\cellcolor[rgb]{0.741,1,0.792}}2,1          & 0,9                                              & 1,0                                               & 0,406                         & 1,685  & 2,497                                               & Stimulation                                    \\ 
\hline
19                                                        & {\cellcolor[rgb]{0.741,1,0.792}}2,2          & 0,9                                              & 1,0                                               & 0,400                         & 1,782  & 2,582                                               & Dependability                                  \\ 
\hline
20                                                        & {\cellcolor[rgb]{0.741,1,0.792}}2,2          & 0,6                                              & 0,8                                               & 0,314                         & 1,913  & 2,541                                               & Efficiency                                     \\ 
\hline
21                                                        & {\cellcolor[rgb]{0.741,1,0.792}}1,2          & 1,2                                              & 1,1                                               & 0,464                         & 0,764  & 1,691                                               & Perspicuity                                    \\ 
\hline
22                                                        & {\cellcolor[rgb]{0.741,1,0.792}}2,0          & 0,7                                              & 0,8                                               & 0,341                         & 1,659  & 2,341                                               & Efficiency                                     \\ 
\hline
23                                                        & {\cellcolor[rgb]{0.741,1,0.792}}1,9          & 1,2                                              & 1,1                                               & 0,452                         & 1,411  & 2,316                                               & Efficiency                                     \\ 
\hline
24                                                        & {\cellcolor[rgb]{0.741,1,0.792}}2,0          & 0,9                                              & 0,9                                               & 0,387                         & 1,613  & 2,387                                               & Attractiveness                                 \\ 
\hline
25                                                        & {\cellcolor[rgb]{0.741,1,0.792}}1,7          & 1,6                                              & 1,2                                               & 0,522                         & 1,160  & 2,204                                               & Attractiveness                                 \\ 
\hline
26                                                        & {\cellcolor[rgb]{0.741,1,0.792}}1,2          & 2,0                                              & 1,4                                               & 0,590                         & 0,637  & 1,817                                               & Novelty                                        \\
\hline
\end{tabular}
}
\end{table}

\begin{table}[h!]
 \centering
 \caption{\label{tab:scaleItems}Mean for each scale in the questionnaire. Green means positive evaluation, and yellow neutral.}
 \scalebox{0.75}{
 \begin{tabular}{|l|c|c|c|c|c|c|} 
 \hline
 \multicolumn{7}{|c|}{{\cellcolor[rgb]{0.502,0.502,0.502}}\textcolor{white}{Confidence intervals (p=0.05) per scale}} \\ 
 \hline
 \rowcolor[rgb]{0.502,0.502,0.502} \textcolor{white}{Scale} & \textcolor{white}{Mean} & \textcolor{white}{Std. Dev.} & \textcolor{white}{N} & \multicolumn{1}{l|}{\textcolor{white}{Confidence}} & \multicolumn{2}{c|}{\textcolor{white}{Confidence interval}} \\ 
 \hline
 Attractiveness & {\cellcolor[rgb]{0.765,0.996,0.769}}1,720 & 0,764 & 22 & 0,319 & 1,401 & 2,039 \\ 
 \hline
 Perspicuity & {\cellcolor[rgb]{0.765,0.996,0.769}}0,898 & 0,925 & 22 & 0,386 & 0,511 & 1,284 \\ 
 \hline
 Efficiency & {\cellcolor[rgb]{0.765,0.996,0.769}}1,966 & 0,784 & 22 & 0,328 & 1,638 & 2,294 \\ 
 \hline
 Dependability & {\cellcolor[rgb]{0.765,0.996,0.769}}1,750 & 0,724 & 22 & 0,302 & 1,448 & 2,052 \\ 
 \hline
 Stimulation & {\cellcolor[rgb]{0.765,0.996,0.769}}1,739 & 0,792 & 22 & 0,331 & 1,408 & 2,070 \\ 
 \hline
 Novelty & {\cellcolor[rgb]{0.765,0.996,0.769}}1,136 & 0,984 & 22 & 0,411 & 0,725 & 1,548 \\
 \hline
 \end{tabular}
 }
 \end{table}

Finally, the UEQ is complemented by a benchmark built using data from 21175 people involved in 468 studies on different digital products, including business software, web pages, web shops, and social networks. 

The benchmark classifies a product into 5 categories and can be interpreted as follows:
\begin{itemize}
  \item Excellent: in the range of the 10\% best results.
  \item Good: 10\% of the results in the benchmark data set are better and 75\% of the results are worse.
  \item Above average: 25\% of the results in the benchmark are better than the results for the evaluated product, 50\% of the results are worse.
  \item Below average: 50\% of the results in the benchmark are better than the results for the evaluated product, 25\% of the results are worse.
  \item Bad: In the range of the 25\% worst results.
\end{itemize}

\begin{figure}[h!]
  \centering
  \includegraphics[scale=0.35]{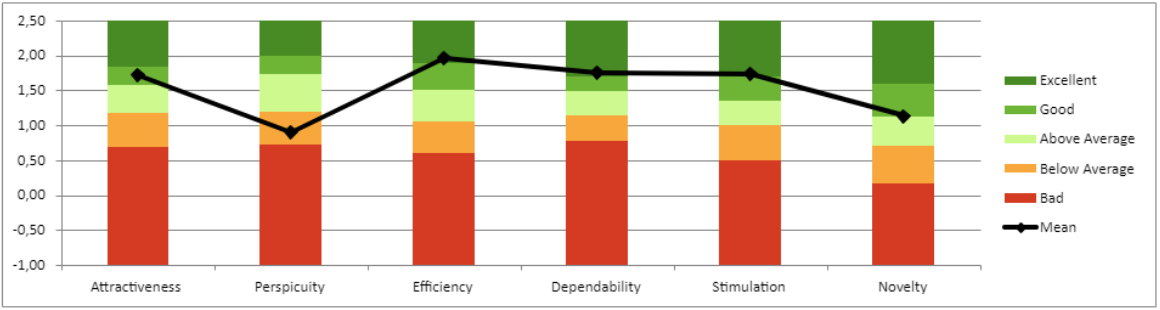}

  \caption{Comparison of SemTUI UEQ results relative to the UEQ benchmark.}
  \label{fig:benchmark}
\end{figure}

In figure \ref{fig:benchmark} the measured scale means are set in relation to the existing values from the benchmark. The results show that SemTUI is positioned in the \textbf{excellent} range for \emph{Efficiency, Dependability} and \emph{Stimulation}, in the \textbf{good} range for \emph{Attractiveness} and \emph{Novelty}, and \textbf{below average} only for \emph{Perspicuity}.

\section{Conclusions and Future Works}
In this paper, we explored the current advances in methods and tools for semantic enrichment of tabular data with the aim of gathering requirements for designing a framework with a graphical interface capable of supporting expert and non-expert users in semantic enrichment tasks.

The evidence from the literature review is that the results of any automated reconciliation tool are far from being completely reliable (i.e., correct and complete), therefore, a human-in-the-loop approach to support enrichment is highly advisable. Moreover, end users should not be exposed to highly specialised tools that require in-depth knowledge of semantic languages and techniques. The developed framework, SemTUI, has a user interface that can effectively engage non-expert users by suggesting reconciliation and extension tools that can be used without specialised knowledge.
A peculiarity of SemTUI is the introduction of a refinement task, a user-centred activity that combines both human knowledge and interactivity to review and perfect the (possibly) ambiguous results of automatic algorithms, improving the semantic annotation and extension with external data.

The evaluation tests confirmed the foreseen usability and effectiveness of the developed user interface. Overall, the product left a good impression and resulted innovative and creative. The comparison of the obtained results from the questionnaire with the usability benchmark of the UEQ method highlighted an area of improvement mainly related to making it more user-friendly and easier to understand and learn, especially for non-expert users.

The architecture of the framework behind the user interface has been carefully designed to provide mechanisms and a solid foundation to support the seamless integration of new services. SemTUI is a modular framework with components, named \emph{Transformers}, devoted to interface external services that can be duplicated, configured, and replaced to simplify the maintenance and evolution of activities and tasks.

A remaining challenge is scalability when applying the enrichment process to large datasets and tables. The tasks of reconciliation and extensions applied on a small batch of data could be reapplied in the future when the same data structure is encountered again. In general, actions performed on a table could be recorded and exported to be reapplied, or to be executed externally as faster scripts on large datasets.

\bibliographystyle{unsrt}  
\bibliography{references}

\begin{thebibliography}{10}

\bibitem{brown2020language}
Tom Brown, Benjamin Mann, Nick Ryder, Melanie Subbiah, Jared~D Kaplan, Prafulla
  Dhariwal, Arvind Neelakantan, Pranav Shyam, Girish Sastry, Amanda Askell,
  et~al.
\newblock Language models are few-shot learners.
\newblock {\em Advances in neural information processing systems},
  33:1877--1901, 2020.

\bibitem{pmlr-v97-brutzkus19b}
Alon Brutzkus and Amir Globerson.
\newblock Why do larger models generalize better? {A} theoretical perspective
  via the {XOR} problem.
\newblock In Kamalika Chaudhuri and Ruslan Salakhutdinov, editors, {\em
  Proceedings of the 36th International Conference on Machine Learning},
  volume~97 of {\em Proceedings of Machine Learning Research}. PMLR, 09--15 Jun
  2019.

\bibitem{azad2018business}
Salahuddin~A Azad, Saleh Wasimi, and ABM~Shawkat Ali.
\newblock Business data enrichment: Issues and challenges.
\newblock In {\em 2018 5th Asia-Pacific World Congress on Computer Science and
  Engineering (APWC on CSE)}, pages 98--102. IEEE, 2018.

\bibitem{jimenez2020semtab}
Ernesto Jim{\'e}nez-Ruiz, Oktie Hassanzadeh, Vasilis Efthymiou, Jiaoyan Chen,
  and Kavitha Srinivas.
\newblock Semtab 2019: Resources to benchmark tabular data to knowledge graph
  matching systems.
\newblock In {\em European Semantic Web Conference}, pages 514--530. Springer,
  2020.

\bibitem{bizer2009emerging}
Christian Bizer.
\newblock The emerging web of linked data.
\newblock {\em IEEE intelligent systems}, 24(5):87--92, 2009.

\bibitem{roman2018datagraft}
Dumitru Roman, Nikolay Nikolov, Antoine Putlier, Dina Sukhobok, Brian
  Elves{\ae}ter, Arne Berre, Xianglin Ye, Marin Dimitrov, Alex Simov, Momchill
  Zarev, et~al.
\newblock Datagraft: One-stop-shop for open data management.
\newblock {\em Semantic Web}, 9(4):393--411, 2018.

\bibitem{maurino2019modelling}
Andrea Maurino, Anisa Rula, Bj{\o}rn~Marius von, Mauricio~Soto Gomez, Brian
  Elves{\ae}ter, and Dumitru Roman.
\newblock Modelling and linking company data in the eubusinessgraph platform.
\newblock In {\em Proceedings of the 5th Workshop on Data Science for
  Macro-modeling with Financial and Economic Datasets}, pages 1--6, 2019.

\bibitem{cutrona2019asia}
Vincenzo Cutrona, Michele Ciavotta, Flavio De~Paoli, and Matteo Palmonari.
\newblock Asia: a tool for assisted semantic interpretation and annotation of
  tabular data.
\newblock In {\em ISWC Satellites}, pages 209--212, 2019.

\bibitem{spahiu2016abstat}
Blerina Spahiu, Riccardo Porrini, Matteo Palmonari, Anisa Rula, and Andrea
  Maurino.
\newblock Abstat: ontology-driven linked data summaries with pattern
  minimalization.
\newblock In {\em European Semantic Web Conference}, pages 381--395. Springer,
  2016.

\bibitem{gupta2012karma}
Shubham Gupta, Pedro Szekely, Craig~A Knoblock, Aman Goel, Mohsen Taheriyan,
  and Maria Muslea.
\newblock Karma: A system for mapping structured sources into the semantic web.
\newblock In {\em Extended Semantic Web Conference}, pages 430--434. Springer,
  2012.

\bibitem{knap2017towards}
Tom{\'a}s Knap.
\newblock Towards odalic, a semantic table interpretation tool in the adequate
  project.
\newblock In {\em LD4IE@ ISWC}, pages 26--37, 2017.

\bibitem{zhang2017effective}
Ziqi Zhang.
\newblock Effective and efficient semantic table interpretation using
  tableminer+.
\newblock {\em Semantic Web}, 8(6):921--957, 2017.

\bibitem{stan}
Riccardo~Porrini Matteo~Palmonari.
\newblock Stan: an incremental semantic annotation tool for tables on the web.
\newblock 2016.

\bibitem{mazumdar2016tool}
Suvodeep Mazumdar and Ziqi Zhang.
\newblock A tool for creating and visualizing semantic annotations on
  relational tables.
\newblock CEUR Workshop Proceedings, 2016.

\bibitem{liu2019dagobah}
Jixiong Liu and Rapha{\"e}l Troncy.
\newblock Dagobah: an end-to-end context-free tabular data semantic annotation
  system.
\newblock {\em SemTab@ ISWC}, 2019.

\bibitem{ongenae2021magic}
Femke Ongenae.
\newblock Magic: Mining an augmented graph using ink, starting from a csv.
\newblock 2021.

\bibitem{steenwinckel2022ink}
Bram Steenwinckel, Gilles Vandewiele, Michael Weyns, Terencio Agozzino,
  Filip~De Turck, and Femke Ongenae.
\newblock Ink: knowledge graph embeddings for node classification.
\newblock {\em Data Mining and Knowledge Discovery}, pages 1--48, 2022.

\bibitem{nguyen2021demonstration}
Phuc Nguyen, Ikuya Yamada, Natthawut Kertkeidkachorn, Ryutaro Ichise, and
  Hideaki Takeda.
\newblock Demonstration of mtab: Tabular data annotation with knowledge graphs.
\newblock 2021.

\bibitem{Cremaschi2020}
Marco Cremaschi, Flavio~De Paoli, Anisa Rula, and Blerina Spahiu.
\newblock A fully automated approach to a complete semantic table
  interpretation.
\newblock {\em Future Gener. Comput. Syst.}, 112:478--500, 2020.

\bibitem{avogadro2021mantistable}
Roberto Avogadro and Marco Cremaschi.
\newblock Mantistable v: A novel and efficient approach to semantic table
  interpretation.
\newblock {\em Semantic Web Challenge on Tabular Data to Knowledge Graph
  Matching (SemTab). CEUR-WS. org}, 2021.

\bibitem{yang2021gbmtab}
Lianzheng Yang, Shuyang Shen, Jingyi Ding, and Jiahui Jin.
\newblock Gbmtab: A graph-based method for interpreting noisy semantic table to
  knowledge graph.
\newblock 2021.

\bibitem{abdelmageed2021jentab}
Nora Abdelmageed and Sirko Schindler.
\newblock Jentab meets semtab 2021’s new challenges.
\newblock {\em Semantic Web Challenge on Tabular Data to Knowledge Graph
  Matching (SemTab). CEURWS. org}, 2021.

\bibitem{baazouzi2021kepler}
Wiem Baazouzi, Marouen Kachroudi, and Sami Faiz.
\newblock Kepler-asi at semtab 2021.
\newblock {\em Semantic Web Challenge on Tabular Data to Knowledge Graph
  Matching (SemTab). CEUR-WS. org}, 2021.

\bibitem{laugwitz2008construction}
Bettina Laugwitz, Theo Held, and Martin Schrepp.
\newblock Construction and evaluation of a user experience questionnaire.
\newblock In {\em Symposium of the Austrian HCI and usability engineering
  group}, pages 63--76. Springer, 2008.

\end{thebibliography}

\end{document}